\documentclass{article}
\usepackage{ijcai25}

\usepackage{times}
\usepackage{soul}
\usepackage{url}
\usepackage[hidelinks]{hyperref}
\usepackage[utf8]{inputenc}
\usepackage[small]{caption}
\usepackage{graphicx}
\usepackage{amsmath}
\usepackage{amsthm}
\usepackage{booktabs}
\usepackage{algorithm}
\usepackage{algorithmic}
\usepackage[switch]{lineno}

\usepackage{color}
\usepackage{xcolor}  
\usepackage{amssymb}
\usepackage{subfigure}
\usepackage{multirow}
\usepackage{pifont}
\usepackage{colortbl}  
\usepackage{natbib}
\usepackage[framemethod=TikZ]{mdframed}

\newcommand\our{\text{TempoLog}} 
\newcommand{\cmark}{\ding{52}}%
\definecolor{dartmouthgreen}{rgb}{0.05, 0.5, 0.06}

\title{Beyond Window-Based Detection: A Graph-Centric Framework for Discrete Log Anomaly Detection}

\author{
Jiaxing Qi\and
Chang Zeng\and
Zhongzhi Luan\and
Shaohan Huang\and
Shu Yang\and
Yao Lu\and
Hailong Yang\And
Depei Qian\\
\affiliations
Sino-German Joint Software Institute, Beihang University, Beijing, China
}

\begin{document}

\maketitle



\begin{abstract}

Detecting anomalies in discrete event logs is critical for ensuring system reliability, security, and efficiency. Traditional window-based methods for log anomaly detection often suffer from \textit{context bias} and \textit{fuzzy localization}, which hinder their ability to precisely and efficiently identify anomalies. To address these challenges, we propose a graph-centric framework, \textbf{\our{}}, which leverages multi-scale temporal graph networks for discrete log anomaly detection. Unlike conventional methods, \our{} constructs continuous-time dynamic graphs directly from event logs, eliminating the need for fixed-size window grouping. By representing log templates as nodes and their temporal relationships as edges, the framework dynamically captures both local and global dependencies across multiple temporal scales. Additionally, a semantic-aware model enhances detection by incorporating rich contextual information. Extensive experiments on public datasets demonstrate that our method achieves state-of-the-art performance in event-level anomaly detection, significantly outperforming existing approaches in both accuracy and efficiency. 
\end{abstract}

\section{Introduction}
Anomaly detection in discrete event sequences plays a crucial role in ensuring modern systems' reliability, security, and efficiency \cite{yuan2021recompose,wang2021multi}. A discrete event sequence is defined as an ordered occurrence of events over time, that are ubiquitous in various domains, including cybersecurity, financial fraud detection, and industrial process monitoring \cite{wang2022maddc,zhang2024multivariate}. By identifying deviations from expected patterns, anomaly detection in these sequences enables timely intervention to prevent potential failures, reduce downtime, and mitigate risks. The inherent complexity of discrete event sequences—characterized by dependencies, temporal relationships, and semantic information—poses significant challenges for detecting anomalies \cite{de2022review}.
\par
\begin{figure}[!tb]
    \centering
    \includegraphics[scale=0.21]{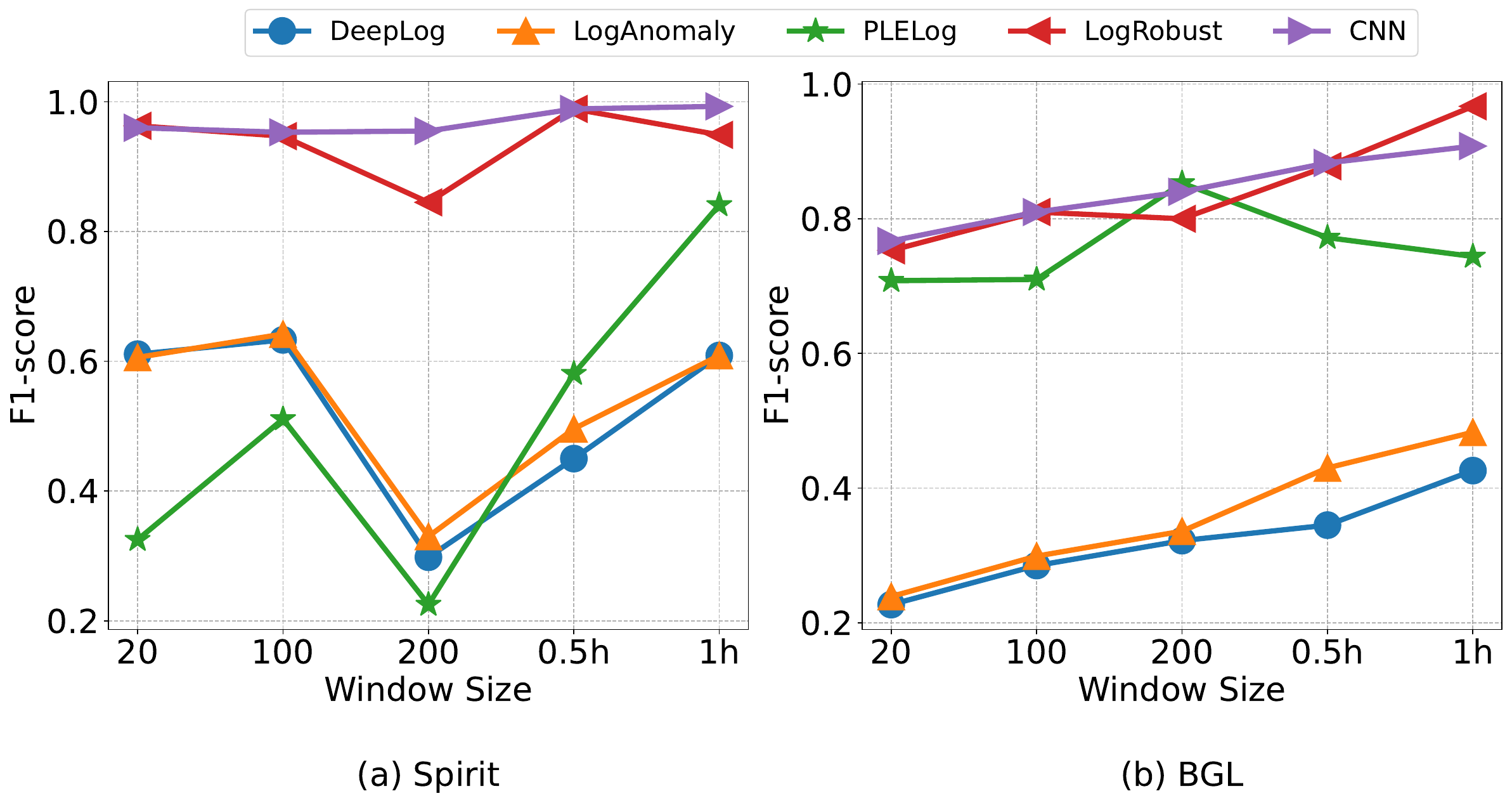}
    \caption{The results for different window sizes—20 logs, 100 logs, 200 logs, 0.5h, and 1h—demonstrate that the performance of various methods is significantly influenced by the choice of window size. These results are referenced from \cite{le2022log}.}
    \label{fig:window_size}
\end{figure}

An example of discrete event sequences is system logs, which record detailed traces of system activities and behaviors.  However, the sheer volume and heterogeneity of logs make manual analysis infeasible. While existing methods, such as RNN-based \cite{landauer2024critical} and Transformer-based methods \cite{catillo2022autolog}, leverage sequential and quantitative patterns in log data, they often rely on fixed-size window grouping to detect anomalies. Consequently, these methods suffer from \textit{context bias} challenge, where the choice of window size significantly impacts detection accuracy \cite{le2022log}. A window that is too small may fail to capture the full context of an anomaly, while a window that is too large may forget critical history information, leading to high false negatives or high false positives. For example, we show the performance of five popular window-based methods, including DeepLog \cite{du2017deeplog}, LogAnomaly \cite{meng2019loganomaly}, PLELog \cite{yang2021semi}, LogRobust \cite{zhang2019robust}, and CNN \cite{lu2018detecting}, on the Spirit and BGL datasets in Figure~\ref{fig:window_size}. The window size significantly affected the final performance. Moreover, another challenge is \textit{fuzzy localization}, as these methods primarily identify anomalies at the window-level rather than the event-level. This means that even when an anomaly is detected, on-site engineers must manually search through the grouped events within the window to locate the certain anomaly event \cite{huang2024demystifying}. This process is time-consuming and significantly undermines efficiency, particularly in large-scale systems handling high-frequency event sequences.

\begin{figure*}
    \centering
    \includegraphics[scale=0.37]{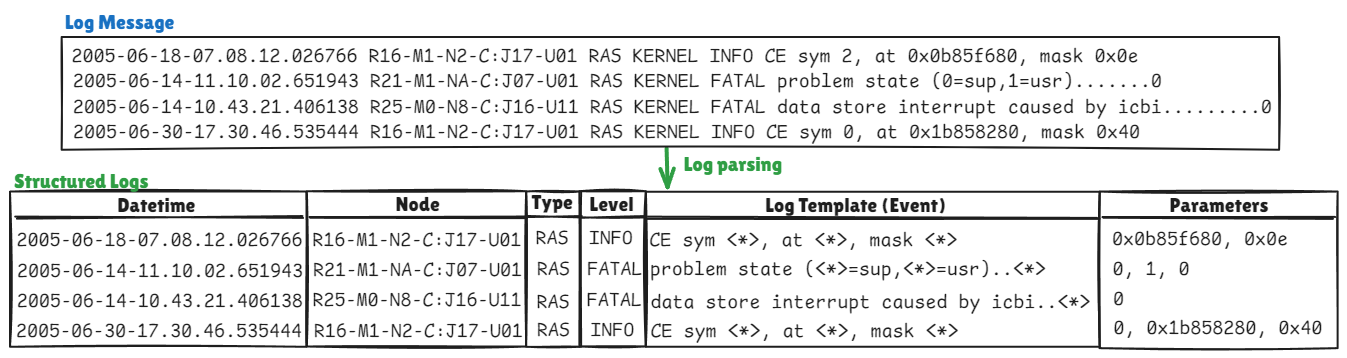}
    \caption{Use log parsing to convert semi-structured log messages into structured logs.}
    \label{fig:parse}
\end{figure*}

\par
Addressing the above challenges requires a paradigm shift from window-based anomaly detection to methods capable of capturing event-level. To this end, we propose \our{}, a multi-scale temporal graph network for discrete event sequence anomaly detection. It is the first to use multi-scale temporal graph for log data, rather than grouping logs into windows. Through the carefully designed graph construction method, log sequences are represented as node events and edge events with timestamps. We then train a multi-scale temporal graph network to capture different hop of event patterns while maintaining efficient parallel processing. Finally, we determine whether an anomaly occurs by predicting the links between two nodes at a specified timestamp. The main contributions of this work are summarized as follows:
\begin{itemize}
    \item To the best of our knowledge, we are the first to introduce multi-scale temporal graph for representing log sequences directly, rather than grouping them into windows, thereby addressing the challenge of \textit{context bias}.
    \item We design a semantic-aware \our{} model that incorporates semantic information to determine the presence of links between nodes at certain timestamps that identify event-level anomalies.
    \item We conduct extensive experiments on three datasets to consistently demonstrate that \our{} achieve state-of-the-art results while ensuring efficiency in both the training and testing phases.
\end{itemize}

\begin{figure*}
    \centering
    \includegraphics[scale=0.3]{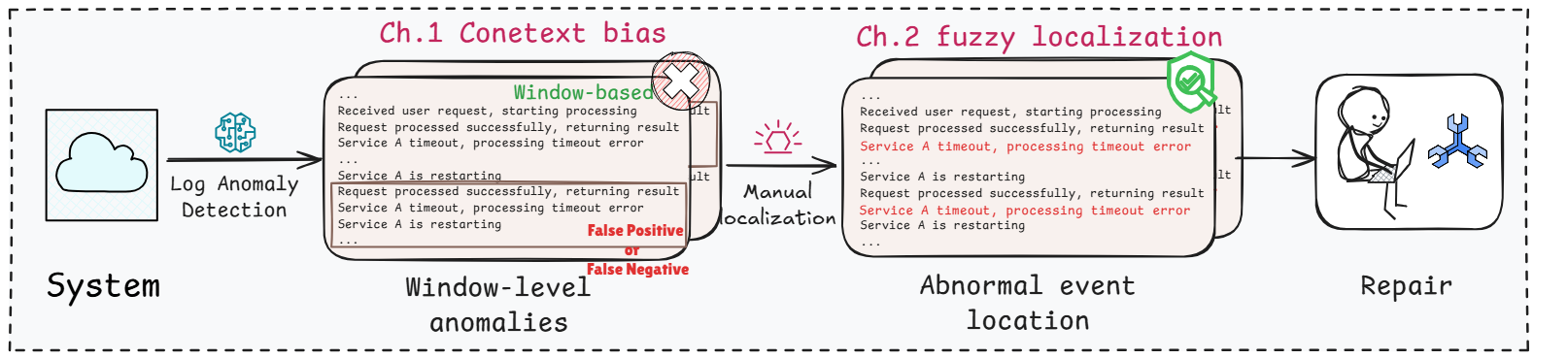}
    \caption{The workflow and challenges of typical window-based log anomaly detection. (DeepLog, LogAnomaly, LogRobust, LogCNN, LogBert, NeuralLog, PLELog, LogGD, and LogGC, etc.)}
    \label{fig:motivation}
\end{figure*}

\section{Background and Motivation}

\subsection{Discrete Log Anomaly Detection}
System logs are typically embedded by developers utilizing logging frameworks like ``\textit{Log4j}" to monitor runtime behavior \cite{hiesgen2022race}. Log messages are semi-structured, a common preprocessing step is formatting each log message into a constant part (also known as ``log template") and a variable part (also known as ``parameters") using log parse techniques. As shown in Figure \ref{fig:parse}, four log messages generate four log templates. Note that the first and second log messages share the same template, ``\textit{CE sym $<$*$>$, at $<$*$>$, mask $<$*$>$}," with only the parameters differing. Log parsing is beyond the scope of this work; therefore, we adopt Drain \cite{he2017drain}, a widely used log parsing method in previous research \cite{le2022log}. Log templates are arranged chronologically based on timestamps to generate a discrete event sequence that reflects the system's running state. Subsequently, deep learning models are employed to detect anomalies within the sequence. Existing methods are categorized into two types based on their model architecture. The first type is sequence-based models, such as DeepLog \cite{du2017deeplog}, while the second type is graph-based models, such as Log2Graph \cite{li2024graph}, LogGD \cite{xie2022loggd}, and LogGC \cite{andonov2023loggc}. Both types require grouping the event sequence into subsequences using a sliding window to detect anomalies at the window-level. The former methods employ RNNs or Transformers to capture sequential patterns, while the latter transform subsequences into graphs and utilize GNNs to learn topological relationships. If an anomaly is identified within a subsequence, the entire subsequence is considered anomalous. Hereafter, we refer to these methods as window-based methods.

\subsection{Motivation}
As shown in Figure \ref{fig:motivation}, during system operation, \textbf{service A} quickly enters an error state after startup, repeatedly encountering timeout errors. The anomaly detection model groups the logs into different windows and detects anomalies within each window. However, the choice of window size may introduce \textit{context bias}. For example, the log message \textit{``Service A timeout, processing timeout error"} represents a true anomaly. Yet, due to insufficient context within the window, the model may incorrectly classify it as a normal timeout error, failing to trigger an alert. Furthermore, even when an anomaly is detected within a window, the \textit{fuzzy localization} of the anomaly requires the on-site engineer to manually pinpoint the location and repair it based on experience. This process is highly time-consuming in practice. Therefore, this work aims to overcome the limitations of window-based methods by proposing a novel solution that eliminates the need for grouping and reduces the workload for on-site engineers.

\section{Design of \our{}}
\subsection{Overview}
\begin{figure*}[!tb]
    \centering
    \includegraphics[scale=0.2]{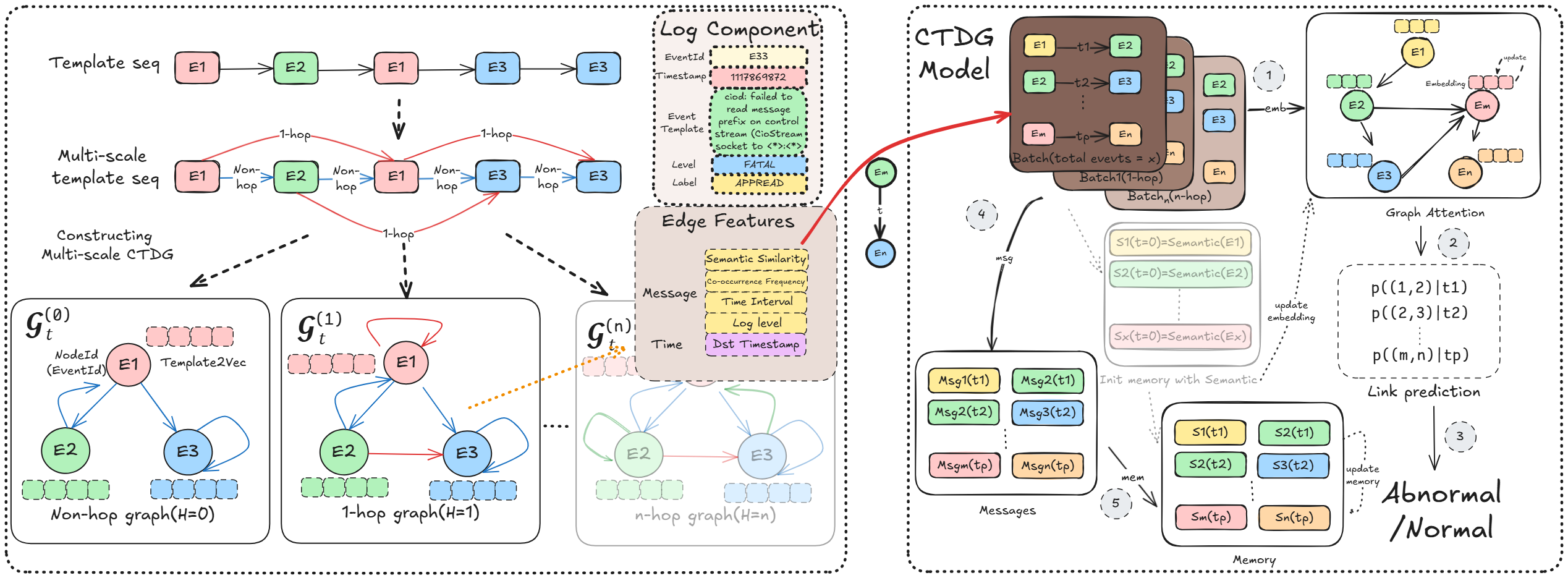}
    \caption{Overview of \our{}. The left panel shows the example of constructing a multi-scale CTDG graph. Each node in the graph is a semantic vector of a log template with a timestamp. The edge features include \textit{semantic similarity}, \textit{co-occurrence frequency}, \textit{time-interval}, and \textit{log level}. The timestamp of the edge is defined as the timestamp of the destination node. The right panel shows event-level anomaly detection using the \our{} model. We leverage a link prediction model to predict whether there is an edge between two nodes at a specific time. If not, an anomaly is detected. At $t=0$, we initialize the memory with the semantic vector of the corresponding log template.}
    \label{fig:overview}
\end{figure*}

In this work, we propose a novel framework, \our{}, that shifts the paradigm from traditional window-based methods to a more flexible and dynamic graph-based representation for modeling discrete event sequences. Our approach uses continuous-time dynamic graphs (CTDGs), which serve as general temporal graph data representations designed to model discrete log sequences over time. In a CTDG, events are represented as a series of nodes and edges that evolve over time, capturing complex relationships such as the addition or deletion of nodes/edges, as well as changes in node and edge features. The design of \our{} is illustrated in Figure \ref{fig:overview}.
The raw log messages are formatted to log templates and represent each template as a dense semantic vector. Then, we construct a multi-scale CTDG based on varying hops of template sequence. Subsequently, we design a semantic-aware \our{} model to represent the template feature at certain timestamps. Finally, a link prediction model determines whether an edge exists between two nodes, enabling end-to-end training.
\par
By leveraging CTDGs, we eliminate the need for fixed-size windows, enabling the model to capture temporal and contextual dependencies between events dynamically. Additionally, the semantic-aware \our{} model ensures event-level anomaly detection by learning patterns across varying hops of event interactions. This design directly addresses \textit{context bias} and \textit{fuzzy localization}, providing a more accurate, interpretable, and efficient solution for anomaly detection in discrete event sequences. 

\subsection{Template2Vec}
Given a raw log sequence $L = \{l_1, l_2, \cdots, l_n\}$, we use Drain to parse the logs into a sequence of log templates $E = \{e_1, e_2, \cdots, e_n\}$. Each log template $e_i$ is a structured representation of the original log message, capturing the essential components such as the timestamp, log level, message type, and other relevant information.
\par
To leverage deep learning models, each log template $e_i$ is represented by a real-valued vector. Pre-trained language models (PLM) like BERT \cite{guo2021logbert} have proven effective in capturing the semantic meaning of text and converting it into dense vectors. Thus, we use BERT to extract the semantic vector for each log template $e_i$.
\begin{equation}
    \begin{array}{cc}
         &  V = PLM(E) = \{ v_1, v_2, \cdots, v_n \} \\
    \end{array}
\end{equation}
where $v_i \in \mathbb{R}^d$ is the semantic vector corresponding to log template $e_i$, and $d$ is the dimensionality of the vector space. 

\subsection{Constructing Multi-scale CTDG}
Next, we model the log sequence as a multi-scale CTDG, where each log template is represented as a node, and the temporal dependencies between log templates are modeled as edges in a dynamic graph. The multi-scale aspect refers to varying the temporal hops (i.e., the number of time steps) considered when defining edges between nodes, which allows us to capture both local dependencies (short hops) and global dependencies (long hops) between log templates. 
\par
Let $\mathcal{G}_t = (\mathcal{V}_t, \mathcal{E}_t)$ denote the CTDG at time $t$, where $\mathcal{V}_t = \{v_1,v_2,\cdots, v_m \}$ is the set of nodes at time $t$. $\mathcal{E}_t$  is the set of directed edges at time $t$, where each edge represents a temporal dependency between log templates (nodes) at consecutive time steps.
\par
To capture both local and global dependencies across log templates, we construct a multi-scale CTDG, which considers different hop distances for temporal edges. Specifically, the multi-scale CTDG is constructed by varying the number of hops between templates when creating edges. This allows the graph to encode different temporal granularities:
\begin{itemize}
    \item \textbf{Local Dependencies}: Short-hop edges between consecutive log templates, capturing immediate temporal relationships.
    \item \textbf{Global Dependencies}: Long-hop edges that capture dependencies over more extended periods, facilitating the understanding of long-term template patterns.
\end{itemize}
\par
Let $H$ represent the hop distance, and define the multi-scale CTDG as a series of graphs over different hops:
\begin{equation}
    \begin{array}{cc}
         &  \mathcal{G}_t^{(H)} = (\mathcal{V}_t^{(H)}, \mathcal{E}_t^{(H)}) ~~ \text{for} ~~ H \in \{0, 1, 2, \cdots\}  \\
    \end{array}
\end{equation}
where $\mathcal{V}_t^{(H)}$ represents the set of nodes at time $t$ for the hop scale $H$, and $\mathcal{E}_t$ represents the set of edges for the same hop scale. The edges capture the dependencies between log templates within an interval of $H$.
\par
Each node $v_i$  in the multi-scale CTDG corresponds to a semantic vector derived from the BERT embedding of log template $e_i$, and edges between these nodes are determined based on the temporal sequence of log templates. For example:
\begin{itemize}
    \item At $H=1$, the edges represent immediate temporal dependencies, such as the occurrence of one log template immediately following another.
    \item At $H>1$, the edges represent long-term dependencies, capturing the relationships between log templates that are further apart in time but may exhibit meaningful patterns over longer sequences.
\end{itemize}
\par
\textbf{Edge features.} To enrich the representation of relationships between log templates, we incorporate the following carefully designed edge features:
\par
\textit{Semantic Similarity (SS):}  Measures the cosine similarity between the semantic vectors of two log templates, indicating how similar their content is.
\begin{equation}
    \begin{array}{cc}
         &  SS(v_i, v_j) = \frac{v_i \cdot v_j}{|| v_i ||~|| v_j||}\\
    \end{array}
\end{equation}
\par
\textit{Co-occurrence Frequency (CF):} Represents how frequently two log templates appear together, capturing their joint occurrence patterns.
\begin{equation}
    \begin{array}{cc}
         &  CF(v_i, v_j) = \frac{\text{Number of co-occurrences of $v_i$ and $v_j$}}{\text{Total number of observations}}
    \end{array}
\end{equation}
\par
\textit{Time Interval (TI):} The time difference between the occurrences of two log templates, provides insight into the temporal gap between events. We noticed that the longer the interval, the greater the likelihood of anomaly. 
\begin{equation}
    \begin{array}{cc}
         &  TI(v_i, v_j) = |timestamp_i - timestamp_j|\\
    \end{array}
\end{equation}
\par
\textit{Log Level (LL):} The log level of a log message (e.g., INFO, WARN, ERROR) is a useful feature for anomaly detection, as anomalous events typically have higher log levels (e.g., ERROR or FATAL). These levels can be represented numerically or encoded as one-hot vectors.
\begin{equation}
    \begin{array}{cc}
         &  LL(v_i) = \text{OneHot(Level of $v_i$)}\\
    \end{array}
\end{equation}

\subsection{Semantic-aware \our{} Model}
For the semantic-aware \our{} model, a CTDG is represented as a sequence of time-stamped events and producing, for each time $t$, the embedding of the graph nodes $Z(t) = (z_1(t), \cdots, z_n(t))$. It includes four core modules: memory, message function, message aggregator, memory updater, and node embedding.    
\par
Each node $v_i$ maintains a memory vector $s_i(t)$ at each timestamp $t$, which is updated based on its previous memory and the incoming messages from its neighbors. Specifically, the message calculation process involves using an \textit{identity} function as the message function, where the message is simply the memory of the neighboring node. The message aggregator then selects the \textit{most recent message} from the neighbors to update the node's memory. The embedding module is implemented using Temporal Graph Attention (TGA) after a trade-off between accuracy and speed, which generates embeddings by attending to the temporal sequence of node interactions. Please see \cite{rossi2020temporal} for more details, the difference is the memory initialization and memory update process.
\par
\textbf{Memory Initialization.} At the start ($t=0$), each node's memory vector $s_i(0)$ is initialized with the semantic vector derived from the pre-trained BERT model for its corresponding log template. This initialization ensures that each node's memory contains meaningful semantic information about the log template it represents, that is $s_i(0) = v_i$. This initialization is important because the semantic vector provides a dense, pre-trained representation of the log template, which serves as the foundation for the subsequent learning process in the TGN.
\par
\textbf{Memory Update.} At each subsequent timestamp $t > 0$, the memory vector $s_i(t)$ is updated by aggregating information from its neighbors based on the edges that exist in the CTDG at time $t$. The update mechanism in \our{} is based on the message passing process, which is influenced by the edge features (e.g., \textit{semantic similarity, co-occurrence frequency, time interval}, and \textit{log level}) as well as the temporal dynamics of the graph. 
\par
Let $\mathcal{N}_i(t)$ represent the set of neighbors of node $v_i$ at time $t$, and let $e_{ij}$ denote the feature vector for the edge between node $v_i$ and node $v_j$. The memory update can be formally as:
\begin{equation}
    \begin{array}{cc}
         &  s_i = \text{mem}(s_i(t-1), \{s_j(t-1), e_{ij}\}_{j\in\mathcal{N}_i(t)}) \\
    \end{array}
\end{equation}

\begin{table}[!tb]
\centering
\caption{The statistics of datasets (Training set ratio=0.5)}
\label{tab:dataset}
\resizebox{1.0\linewidth}{!}{
\begin{tabular}{cccc} 
\toprule
                   & \textbf{BGL}    & \textbf{Spirit}  & \textbf{Thunderbird}     \\ 
\midrule
\# Logs            & 4,747,963       & 500,000,000      & 9,959,159       \\
\# Template        & 1,847           & 2,880            & 4,992           \\
\# Anomalies       & 348,460(7.39\%) & 764,891(15.30\%) & 4,934(0.05\%)           \\
\# Train logs      & 2,356,746       & 25,000,000       & 4,979,580       \\
\# Train Anomalies & 229,848(4.88\%) & 591,265(11.82\%) & 3,273 (0.03\%)  \\
\# Test Anomalies  & 118,612(2.52\%) & 173,626(3.47\%)  & 1,661(0.02\%)   \\
\bottomrule
\end{tabular}}
\end{table}

\begin{table*}[!tb]
\centering
\caption{Overall performance}
\label{tab:overall}
\begin{tabular}{c|ccc|ccc|ccc} 
\toprule
\textbf{Dataset}   & \multicolumn{3}{c|}{\textbf{BGL}}     & \multicolumn{3}{c|}{\textbf{Spirit}}  & \multicolumn{3}{c}{\textbf{Thunderbird}}  \\ 
\midrule
                   & \multicolumn{3}{c|}{\textbf{Evaluate Metrics}} & \multicolumn{3}{c|}{\textbf{Evaluate Metrics}} & \multicolumn{3}{c}{\textbf{Evaluate Metrics}}      \\ 
\cmidrule{2-10}
\textbf{Algorithm} & F1     & Precision & Recall           & F1     & Precision & Recall           & F1     & Precision & Recall                \\ 
\midrule
DeepLog    & 0.226~ & 0.119~    & 0.954~ & 0.186~ & 0.984~    & 0.103~ & 0.007~ & 0.003~     & 0.947~     \\
LogBert    & 0.920~ & 0.890~    & 0.951~ & 0.975~  & 0.964~ & 0.986~ & 0.405~ & 0.833~ & 0.268           \\
\midrule
CNN        & 0.406~ & 0.267~    & 0.851~ & 0.948~ & 0.958~    & 0.938~ & 0.637~ & 0.855~    & 0.508~      \\
LogAnomaly & 0.224~ & 0.126~    & 0.976~ & 0.193~ & 0.107~    & 0.991~ & 0.007~ & 0.003~    & 0.992~      \\
LogRobust  & 0.383~ & 0.256~    & 0.762~ & 0.610~ & 0.543~    & 0.697~ & 0.175~ & 0.106~    & 0.500~      \\
PLELog     & 0.875~ & 0.847~    & 0.904~ & 0.909~ & 0.986~    & 0.948~ & 0.440~ & 0.998~    & 0.484~       \\ 
NeuralLog  & 0.370~ & 0.239~    & 0.818~ & 0.423~ & 0.234~    & 0.923~ & 0.569~ & 0.660~    & 0.500~      \\ 
\midrule
\our{}  & \cellcolor{lightgray} \textbf{0.986~} & \cellcolor{lightgray} \textbf{0.990~}   & \cellcolor{lightgray} \textbf{0.982~} & \cellcolor{lightgray} \textbf{0.993~} & \cellcolor{lightgray} \textbf{0.997~}    & \cellcolor{lightgray} \textbf{0.988~} & \cellcolor{lightgray} \textbf{0.987~} & \cellcolor{lightgray} \textbf{0.992~}    & \cellcolor{lightgray} \textbf{0.983~}      \\
\our{} \textit{w/o semantic}  &0.728~  & 0.850~     &0.637~  &0.250~  &1.000~     &0.143~  &0.518~  &0.935~     & 0.358~      \\
\our{} ($H=0$) &0.839~  & 0.828~     &0.850~  &0.371~  &0.956~     &0.230~  &0.388~  &0.324~     & 0.483~      \\
\bottomrule
\end{tabular}
\end{table*}

\subsection{Event-level Anomaly Detection}
We employ a link prediction model to determine whether an edge will exist between two nodes at a given timestamp $t$, which can be formulated as:
\begin{equation}
    \begin{array}{cc}
         &   L(\Theta) = \text{BCE}(y_{ij}, \sigma(f(z_i, z_j)))
    \end{array}
\end{equation}
where $\hat{y}_{ij}$ is the predicted probability of the existence of an edge between nodes $v_i$ and $v_j$ at time t, $f$ is the link prediction model, and $\sigma$ is a sigmoid activation function. BCE is the binary cross entropy loss function, $\Theta$ is all trainable parameters. Notably, we train a \our{} model for each graph of different hops. To reduce the training parameters, we share their parameters, which greatly improves the training and inference efficiency. In the detection stage, given a log sequence, if the predicted link relationship is inconsistent with the real link relationship within any hop input graph $\mathcal{G}^{(H)}_t$, an event-level anomaly is detected, otherwise it is normal. 
\par
\textbf{Unseen Log Detection}: For a new log, if the extracted template does not already exist, we extract its semantic vector, increase the memory size by 1, and add the semantic vector to the memory. Additionally, in CTDG, an event of node addition is triggered, and a new node is created using the semantic vector and determines whether it is anomaly or not.

\section{Evaluation}
\subsection{Experiment Setting}
\subsubsection{Datasets} 
We conduct experiments on three public datasets that have been widely used in previous work: BGL, Spirit, and Thunderbird. In the following experiments, we chronologically select 50\% logs as training set, and the rest as testing set. The statistics of all datasets are presented in Table~\ref{tab:dataset}. The anomaly ratios of the three datasets are quite small, ranging from $\sim$0.05\% to $\sim$15\%. Additionally, all datasets contain a large number of templates, with many templates in the test set not appearing in the training set. These factors contribute to the difficulty of the anomaly detection on these datasets.

\subsubsection{Baselines} 
To better demonstrate the performance of \our{}, we include a wide range of state-of-the-art baseline algorithms. The baselines can be divided into two groups: (i) non-semantic information methods that only use log template indices as input (\textbf{DeepLog, LogBert}). (ii) semantic-aware methods that use the semantic vector of each log template/message as input (\textbf{CNN, LogAnomaly, LogRobust, PLELog, NeuralLog}). 

\subsubsection{Evaluation Metrics} 
Anomaly detection is a binary classification task, we use the F1-score, Precision, and Recall metrics to measure the effectiveness of models. We label its outcome as a TP, TN, FP, and FN. TP are the anomalous event that are accurately detected. TN are normal event that are accurately detected. If detect an event as anomalous but normal, we label the outcome as FP and the reset are FN. Precision = $\frac{\text{TP}}{\text{TP}+\text{FP}}$, Recall=$\frac{\text{TP}}{\text{TP}+\text{FN}}$, and F1-score = $\frac{2*\text{Precision} * \text{Recall}}{\text{Precision} + \text{Recall}}$.

\subsubsection{Environment Setup}
Our experiments were conducted within a Docker container based on Ubuntu. The machine used has a CPU consisting of 24 Intel(R) Xeon(R) Gold 6126 CPUs @ 2.60GHz and two Tesla V100S-PCIE-32GB GPUs. For comparison, we utilized methods from the open-source LogADEmpirical framework \footnote{https://github.com/LogIntelligence/LogADEmpirical}. As for LogBert, we use the open-source implementation released in the paper \footnote{https://github.com/HelenGuohx/logbert}.

\subsection{Evaluation of The Overall Performance}
In this section, we compare the performance of \our{} with seven other methods across three datasets. To ensure the fairness of the results, we consistently allocate 40\% of the data as the test set. The processed logs are directly input into \our{}, while for the other window-based models, we utilize a fixed-size window to group logs and the window size is 100. By default, we take $H = \{0, 1\}$, i.e. the 0-hop and 1-hop graphs are used for both training and testing. The results are presented in Table \ref{tab:overall}.
\par
We can see that \our{} demonstrates strong performance across all three datasets. Notably, on the BGL dataset, \our{} achieves an F1-score of 0.986, which is significantly higher than the best-performing LogBert, with an F1-score of 0.920. On the Spirit dataset, \our{}'s F1-score achieves 0.993, surpassing the best LogBert model's F1-score of 0.948. On the Thunderbird dataset, \our{} achieves an F1-score of 0.987, which is notably higher than CNN's 0.637. However, DeepLog and LogAnomaly do not effectively perform on Thunderbird, indicating that this dataset poses considerable challenges. We believe that the anomaly ratio in the Thunderbird dataset is extremely low (0.05\%) and that many templates in the test set are unseen. In addition, both DeepLog and LogAnomaly detect anomalies by predicting the next log template. These factors result in a high false positive rate of the model.
\par
In addition, we also compare \our{} \textit{w/o semantic} and \our{} with only non-hop input, i.e. $H=0$ (the last two rows in Table \ref{tab:overall}). The results show that both \textit{w/o semantic} and only non-hop input significantly decrease the F1-score. On the BGL, Spirt and Thunderbird datasets, the F1-score of \our{} \textit{w/o semantic} dropped by $\sim$26\%, $\sim$~75\% and $\sim$48\% respectively, and that of non-hop \our{} dropped by $\sim$15\%, $\sim$63\% and $\sim$61\% respectively. These results demonstrate that adding semantic information to memory and multi-scale input can improve detection performance.

\subsection{Ablation Study}
We also conducted some ablation experiments to check how each edge feature affects the detection accuracy. The results are presented in Table \ref{tab:edge_features}.
\par
The results demonstrate that all four edge features contribute significantly to the model's performance. On the BGL and Spirit datasets, the F1-scores with all features enabled achieve 0.986 and 0.993, respectively, representing the best performance. Removing any single feature leads to a noticeable loss in performance. For example, removing \textit{log level} (C4) reduces the F1-score to 0.933 on BGL and 0.968 on Spirit. On the BGL dataset, removing C4 slightly increases the false alert rate and yields a higher recall of 0.983, indicating the model detects more true anomaly logs. 
\par
Among the aforementioned features, \textit{semantic similarity} (C1) and \textit{time interval} (C3) have a significant impact on the BGL dataset, as their removal results in the most significant decline in model performance, while the influence of \textit{co-occurrence frequency} (C2) is minimal. In contrast, for the Spirit dataset, the \textit{time interval} has the most substantial effect, whereas \textit{semantic similarity} (C1) has only a negligible impact on the model's performance. In a nutshell, the degree of influence of different features varies across datasets, but the \textit{time interval} (C3) consistently plays a relatively significant role. The absence of any single feature would result in a degradation of the model's performance.


\begin{table}[!tb]
\centering
\caption{Ablation study of edge features. \textbf{C1}: \textit{semantic similarity}, \textbf{C2}: \textit{co-occurrence frequency}, \textbf{C3}: \textit{time interval}, \textbf{C4}: \textit{log level}}
\label{tab:edge_features}
\resizebox{1.0\linewidth}{!}{
\begin{tabular}{cccccccccc} 
\toprule
\multirow{2}{*}{\textbf{C1}} & \multirow{2}{*}{\textbf{C2}}  & \multirow{2}{*}{\textbf{C3}}  & \multirow{2}{*}{\textbf{C4}}  & \multicolumn{3}{c}{\textbf{BGL}} & \multicolumn{3}{c}{\textbf{Spirit}}   \\ 
\cmidrule{5-10}
\multicolumn{4}{c}{}     & F1 & Precison & Recall  & F1 & Precison & Recall      \\ 
\cmidrule{5-10}
& \color{dartmouthgreen} \cmark  & \color{dartmouthgreen} \cmark  &  \color{dartmouthgreen} \cmark               & 0.881~   & 0.869~   & 0.894~  & 0.988~  & 0.993~   & 0.983~       \\
\color{dartmouthgreen} \cmark   &   & \color{dartmouthgreen} \cmark & \color{dartmouthgreen} \cmark              & 0.980~   & 0.880~    & 0.937~  & 0.963~  & 1.000~   & 0.929~             \\
\color{dartmouthgreen} \cmark   & \color{dartmouthgreen} \cmark  &   & \color{dartmouthgreen} \cmark             & 0.898~   & 0.973~    & 0.834~  & 0.921~  & 0.984~   & 0.867~        \\
\color{dartmouthgreen} \cmark   & \color{dartmouthgreen} \cmark  & \color{dartmouthgreen} \cmark  &              & 0.933~   & 0.887~    & 0.983~  & 0.968~   & 0.985~         & 0.952~            \\
\color{dartmouthgreen} \cmark   & \color{dartmouthgreen} \cmark  & \color{dartmouthgreen} \cmark  & \color{dartmouthgreen} \cmark   & 0.986~   & 0.990~    & 0.982~  & 0.993~   & 0.997~         & 0.988~            \\
\bottomrule
\end{tabular}}
\end{table}

\subsection{Case Study}


A common challenge arises when two related log messages are separated by a significant temporal distance. When using fixed-size windows for grouping, these related logs may be assigned to different windows. Traditional window-based detection methods struggle to capture the relationships between such logs due to their separation. As a result, the anomalous information they contain may remain undetected.
\par
\begin{figure}[!tb]
    \centering
    \includegraphics[scale=0.31]{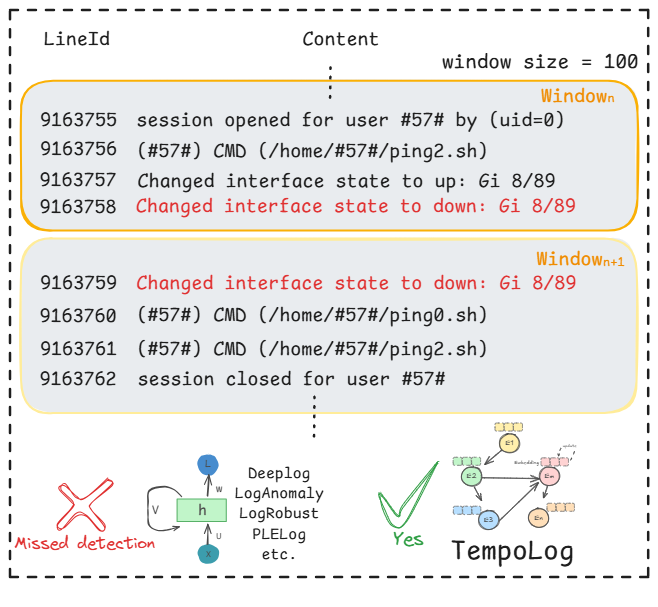}
    \caption{An example of \textit{context bias} and \textit{fuzzy localization} caused by the use of a fixed-size window on the Thunderbird dataset.}
    \label{fig:case}
\end{figure}

An example is illustrated in Figure \ref{fig:case}, which presents a sampled log sequence from the Thunderbird dataset. In this example, a user initiates a session, changes the network interface status to ``up," and subsequently switches it to ``down." However, the log records two consecutive ``down" statuses, indicating a potential system issue. From the perspective of $\text{Window}_n$, \textit{``Gi 8/89"} undergoes a typical transition from ``up" to ``down," which appears normal. In contrast, from the perspective of $\text{Window}_{n+1}$, the prior state of \textit{``Gi 8/89"} is unknown, making it impossible to identify the anomaly based solely on this log message.
While it is possible to set different time step lengths during window segmentation to mitigate \textit{context bias}, if the anomaly logs are distant, the cost of window-based models increases, and the challenge of \textit{fuzzy localization} becomes more pronounced. \our{} capturing event-level relationships within continuous-time dynamic graphs, effectively addresses and detects such challenges.

\begin{figure}
    \centering
    \includegraphics[scale=0.21]{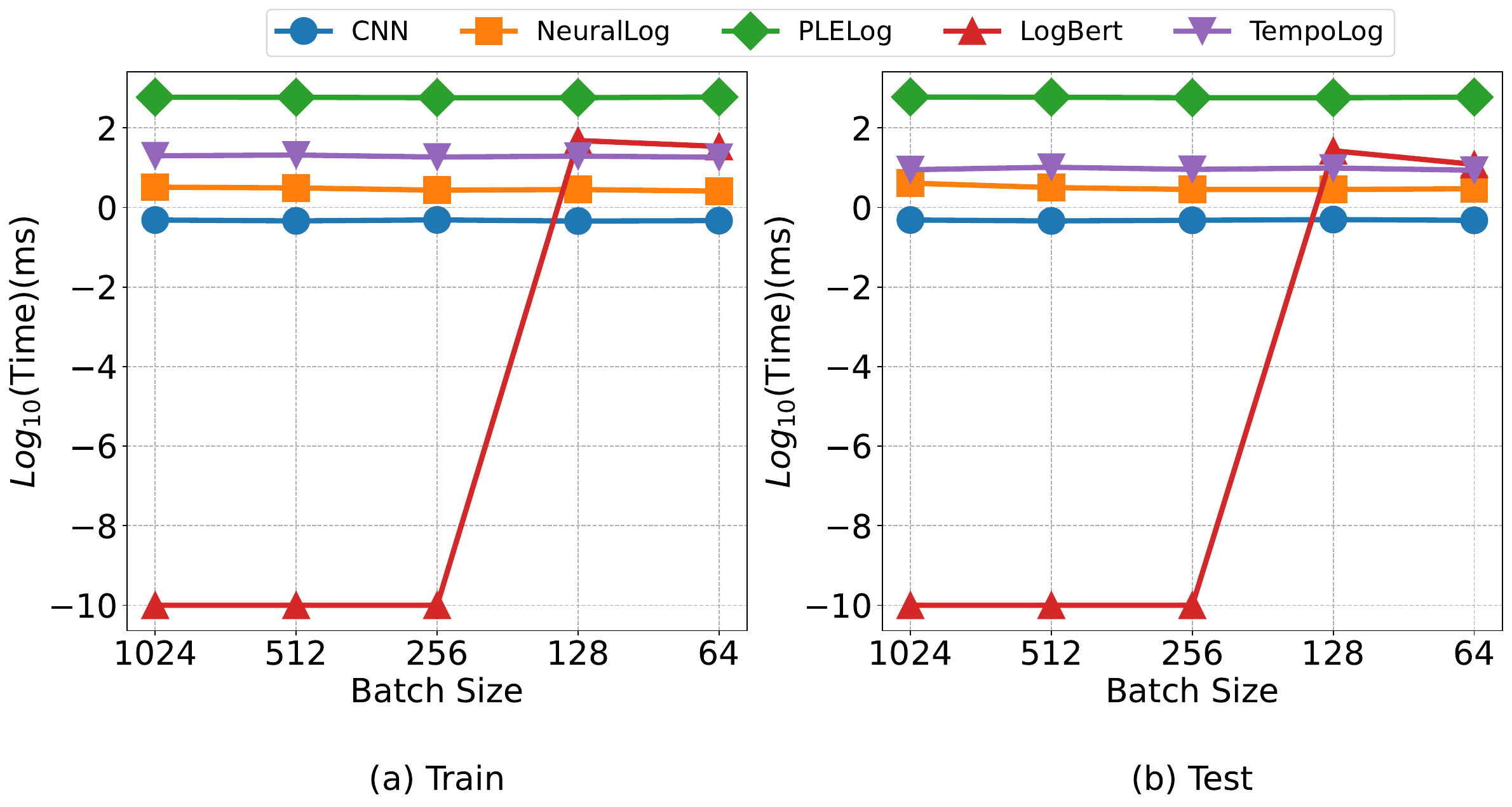}
    \caption{Comparison of training and testing times for different models with batch sizes ranging from 1024 to 64. LogBert encounters out-of-memory with batch sizes of 1024, 512, and 256.}
    \label{fig:time}
\end{figure}

\subsection{Cost and Efficiency}
Practitioners prioritize real-time performance and scalability \cite{ma2024practitioners}. Therefore, we conducted a statistical analysis of the time overhead during the model training and inference stages, as well as GPU memory usage.
\par
We compared the time cost of \our{} and other baseline models in the training and testing phases, as shown in Figure~\ref{fig:time}. The time cost of \our{} is lower than that of PLELog and LogBert due to the higher complexity of these two models. Conversely, the time cost of CNN and NeuralLog is lower than that of \our{}. However, during the test phase, as the batch size increases, the time cost of \our{} gradually approaches that of NeuralLog, highlighting the efficient parallelism of \our{}. Finally, it is worth noting that \our{} detects event-level anomalies, effectively addressing the \textit{fuzzy localization} challenge inherent in window-based methods. Although the time overhead during the testing phase is higher than that of CNN and NeuralLog, \our{} eliminates the need for manual anomaly localization—a benefit that far outweighs the additional time cost. Consequently, this overhead is considered acceptable.

\par
\begin{figure}[!tb]
    \centering
    \subfigure[Spirit]{\includegraphics[scale=0.26]{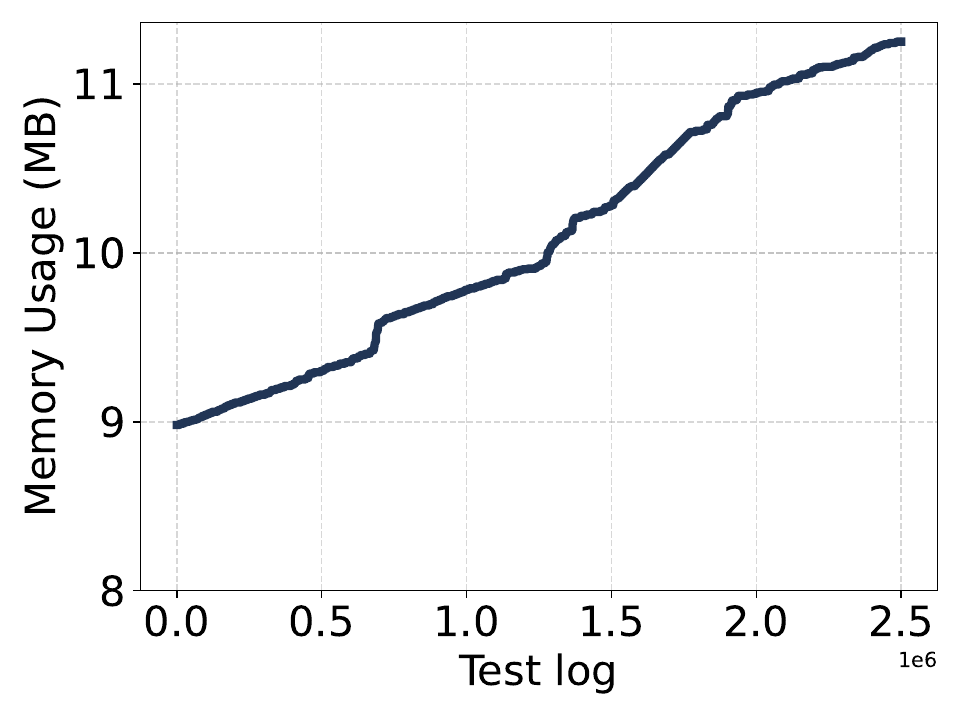}}
    \subfigure[Thunderbird]{\includegraphics[scale=0.26]{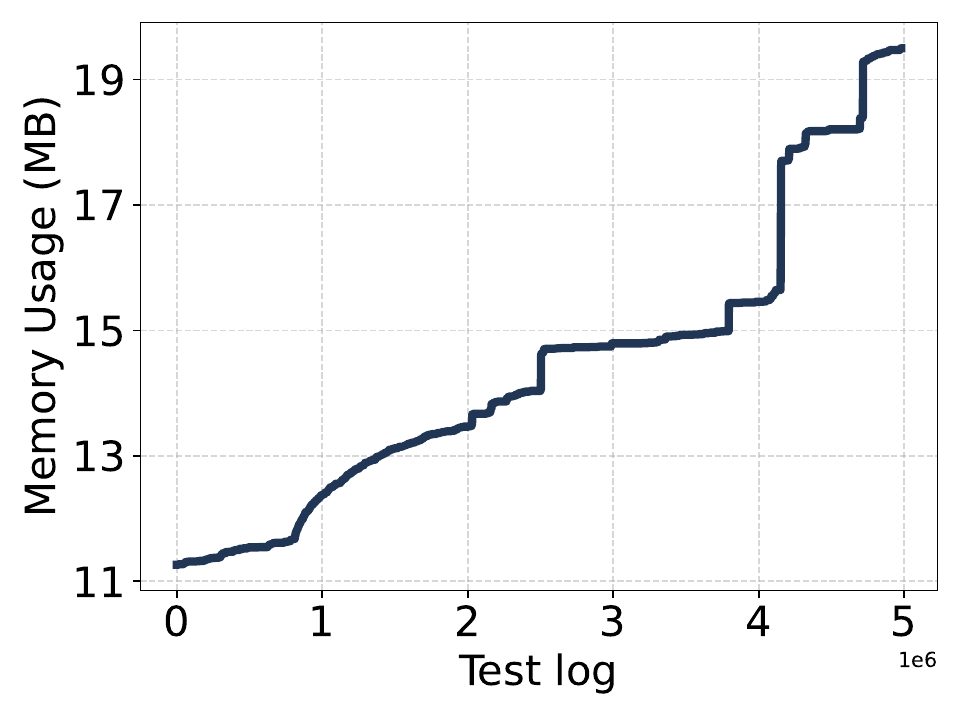}}
    \caption{GPU memory usage of \textit{memory module} in test phase. }
    \label{fig:memory}
\end{figure}

The \textit{memory module} stores the states of all log templates up to the current time $t$. Many templates are unseen during the test phase, so the state information in memory needs to be updated. We analyzed the memory usage of the \textit{memory module} on both Spirit and Thunderbird test sets. As shown in Figure~\ref{fig:memory}, there are a large number of unseen templates in the test sets of both datasets, simulating the scenario of dynamic log updates in practice. We can see that as the number of test logs increases, unseen templates are updated in memory and the GPU memory also increases, but the memory overhead of the \textit{memory module} is kept at a very low value.

\subsection{Threats to Validity}
The following threats to validity may affect our study findings. \textbf{(i) Unstable Training}. We found that performance oscillations occurred during the training of some datasets (such as BGL). This may be attributed to the characteristics of the dataset itself, such as the complexity of the log pattern or noise interference; it may also be due to the model itself, such as unstable parameter initialization or the model's sensitivity to small batches of data. This performance oscillation phenomenon makes it difficult for the model to converge stably during training, which not only prolongs the training time, but also may affect the performance and generalization ability of the final model. \textbf{(ii) Template Explosion.} Different log parsing methods may generate different numbers of log templates. In practical applications, the number of log templates can increase over time and as system complexity increases. These two factors will significantly increase the GPU memory usage of the memory module, especially when processing large log data or long-running systems, which can lead to excessive resource consumption and affect the training efficiency and inference speed.



\section{Conclusion and Future Work}
In this paper, we propose \our{}, a window-free log anomaly detection framework that leverages temporal graph networks to capture the semantic information of log templates at each timestamp and addresses the challenges inherent in window-based methods. The model integrates well-designed node and edge features of multi-scale continuous-time dynamic graphs. Extensive experiments conducted on three log datasets, compared against seven baselines, demonstrate the superior performance of \our{}. The ablation study further validates the effectiveness of the multi-scale input and edge feature design. The case study shows that \our{} not only detects event-level anomalies but also significantly reduces the time required for manual localization.
\par
In future work, we consider improving the training and inference efficiency of the model and exploring more ways to construct continuous-time dynamic graphs.

\bibliographystyle{named}
\bibliography{ijcai25}

\end{document}